\documentclass[preprintnumbers,10pt,nofootinbib]{revtex4}
\pdfoutput=1

\usepackage{amsmath,latexsym,amssymb,amsfonts}
\usepackage[pdftex]{color,graphicx}
\usepackage{bm}
\usepackage{subfigure}

\begin{document}


\title{\textbf{Fuzzy Euclidean wormholes in the inflationary universe}}

\author{
\textsc{Pisin Chen$^{a,b,c,d}$}\footnote{{\tt pisinchen{}@{}phys.ntu.edu.tw}},
\textsc{Daeho Ro$^{e,f}$}\footnote{{\tt daeho.ro{}@{}bespinglobal.com }}
and
\textsc{Dong-han Yeom$^{e,g,h}$}\footnote{{\tt innocent.yeom{}@{}gmail.com}}
}
\affiliation{
$^{a}$\small{Leung Center for Cosmology and Particle Astrophysics, National Taiwan University, Taipei 10617, Taiwan}\\
$^{b}$\small{Department of Physics, National Taiwan University, Taipei 10617, Taiwan}\\
$^{c}$\small{Graduate Institute of Astrophysics, National Taiwan University, Taipei 10617, Taiwan}\\
$^{d}$\small{Kavli Institute for Particle Astrophysics and Cosmology, SLAC National Accelerator Laboratory, Stanford University, Stanford, California 94305, USA}\\
$^{e}$\small{Asia Pacific Center for Theoretical Physics, Pohang 37673, Republic of Korea}\\
$^{f}$\small{Bespin Global, Seoul 06627, Republic of Korea}\\
$^{g}$\small{Department of Physics, POSTECH, Pohang 37673, Republic of Korea}\\
$^{h}$\small{Department of Physics Education, Pusan National University, Busan 46241, Republic of Korea}
}

\begin{abstract}
In this paper, we investigate complex-valued Euclidean wormholes in the Starobinsky inflation. Due to the properties of the concave inflaton potential, the classicality condition at both ends of the wormhole can be satisfied, as long as the initial condition of the inflaton field is such that it is located sufficiently close to the hilltop. We compare the probabilities of classicalized wormholes with the Hartle-Hawking compact instantons and conclude that the Euclidean wormholes are probabilistically preferred than compact instantons, if the inflation lasts more than $50$ e-foldings. Our result assumes that the Euclidean path integral is the correct effective description of quantum gravity. This opens a new window for various future investigations that can be either confirmed or refuted by future experiments.
\end{abstract}

\maketitle

\newpage

\tableofcontents

\section{Introduction}

Understanding the origin of our universe is an important task in cosmology. From recent cosmological observations \cite{Akrami:2018odb}, we now confirm the existence of an inflation epoch in the very beginning of the universe \cite{Guth1981}, which is very well described by a concave inflaton potential examplified by the Starobinsky model \cite{Starobinsky1980}.

There remains, however, several problems. For example, the problems of the initial singularity \cite{DeWitt1967} and the proper initial conditions that give rise to the inflation remain unresolved \cite{Hartle2008a,Hwang2014}. It is reasonable to expect that a future full-blown quantum theory of gravity might be able to address these issues, based on which the observed cosmic microwave background (CMB) parameters would be shown as typical or natural consequence of such initial conditions.

In short of a final quantum gravity theory, in this paper we invoke the Euclidean path-integral as the wave function of the universe \cite{Hartle1983} to adress the issue. The Euclidean path-integral has many nice properties including its satisfying the Wheeler-DeWitt equation \cite{DeWitt1967,Hartle1983}. In addition, if one restricts the path-integral solutions to compact instantons only, that is, if one invokes the no-boundary proposal \cite{Hartle1983}, then the wave function corresponds to the ground state and provides a consistent thermodynamical limit of the stochastic distribution of universes \cite{Linde:1993xx}. On the other hand, this ansatz has a weak point: the wave function exponentially prefers a small number of $e$-foldings of inflation that is in conflict with the observations \cite{Vilenkin:1987kf}.

Because of this reason, there have been some authors have raised doubts about the no-boundary proposal \cite{Vilenkin:1987kf}. With that in mind, there have been attempts to attain large $e$-foldings within the Euclidean path-integral formalism. For example, Hartle, Hawking, and Hertog proposed to include the volume-weighting to the wave function as a means to extend the inflation $e$-foldings \cite{Hartle2008a}. Other attempts include fine-tunings of inflation models \cite{Hwang2014,Hwang:2012bd}, modified gravity \cite{Sasaki:2013aka}, and the introduction of an additional, more massive field \cite{Hwang2015}.

In the present work, we will not specifically choose a certain hypothesis that explains the large $e$-foldings in the Euclidean path-integral approach. Rather, we will demonstrate that, so long as one imposes the restriction that the inflation must be lasted for more than 50 $e$-foldings, the wave function of the universe would prefer non-compact instantons over compact ones.  The latter correspond to instantons proposed by Hartle and Hawking, while the former correspond to the so-called Euclidean wormholes \cite{Chen:2016ask,Chen:2018atc,Kang:2017jmq}. This implies that our universe is more likely to have emerged from a Euclidean wormhole rather than a compact instanton, which in turn would give raise to several interesting implications.

This paper is organized as follows. In Sec.~\ref{sec:pre}, we summarize our previous investigations about Euclidean wormholes. In Sec.~\ref{sec:Euc}, we consider the classicality condition in the Starobinsky inflation model, and calculate the probability distribution. We then compare the probability of wormholes to that of compact instantons. We conclude that, as long as one restricts duration of the inflation be longer than 50 $e$-foldings, then the Euclidean wave function prefers not compact but non-compact instantons. Finally, in Sec.~\ref{sec:con}, we summarize our results and comment on possible future topics.

\section{\label{sec:pre}Preliminaries}

In this section, we summarize previous results about fuzzy Euclidean wormholes in de Sitter space.

\subsection{The Euclidean path-integral approach}

The Euclidean path-integral is described by \cite{Hartle1983}
\begin{eqnarray}
\Psi[h_{\mu\nu}^{\mathrm{f}}, \chi^{\mathrm{f}}; h_{\mu\nu}^{\mathrm{i}}, \chi^{\mathrm{i}}] = \int \mathcal{D}g\mathcal{D}\phi \;\; e^{-S_{\mathrm{E}}[g,\phi]},
\end{eqnarray}
where $S_{\mathrm{E}}$ is the Euclidean action, $g_{\mu\nu}$ is the metric, $\phi$ is a matter field, and we integrate over all geometries that have $h_{\mu\nu}^{\mathrm{i,f}}$ and $\chi^{\mathrm{i,f}}$ as their boundary values (i and f denote initial and final hypersurfaces, respectively). This will be approximated by steepest-descents, or equivalently, the Euclidean instantons. Here, it is important to mention that in order to assign well-defined probabilities, we need to restrict ourselves to dealing only with regular instantons. As a consequence, we need to carefully choose boundary conditions of each instanton to find regular solutions.

In general, the Euclidean instantons have two boundaries (initial and final boundaries), but if the instanton is compact and hence the initial part and the final part are smoothly disconnected from each other, then one can consider a wave function that only has its future boundary defined. This is the origin of the terminology \textit{no-boundary proposal}, since the initial condition of the wave function, $\Psi[h^{\mathrm{f}}_{\mu\nu},\xi^{\mathrm{f}}]$, has \textit{no} boundary. This is, nevertheless, a special and restricted choice; indeed, it is more general to consider non-compact instantons which have two boundaries. The simplest example of this category is the \textit{Euclidean wormhole} \cite{Chen:2016ask,Kang:2017jmq}.

Keeping these two kinds of instantons in mind, let us consider the following action
\begin{eqnarray}
S = \int \sqrt{-g}dx^{4} \left[ \frac{R}{16\pi} - \frac{1}{2} \left(\nabla \phi\right)^{2} - V(\phi) \right].
\end{eqnarray}
In addition, we define the Euclidean minisuperspace metric as follows
\begin{eqnarray}
ds^{2}_{\mathrm{E}} = d\tau^{2} + a^{2}(\tau) d\Omega_{3}^{2},
\end{eqnarray}
where $a(\tau)$ is the scale factor and $d\Omega_{3}^{2}$ is the solid angle of the three-sphere. Then the equations of motion that instantons have to satisfy are as follows:
\begin{eqnarray}
\label{eq:1}\dot{a}^{2} - 1 - \frac{8\pi a^{2}}{3} \left( \frac{\dot{\phi}^{2}}{2} - V \right) &=& 0,\\
\label{eq:2}\ddot{\phi} + 3 \frac{\dot{a}}{a} \dot{\phi} - V' &=& 0, \\
\label{eq:3}\frac{\ddot{a}}{a} + \frac{8\pi}{3}\left( \dot{\phi}^{2} + V \right) &=& 0.
\end{eqnarray}

\subsection{Fuzzy Euclidean wormholes in de Sitter space}

Because of the analyticity condition of the Wick-rotation, we require that all functions be complex-valued \cite{Hwang2013a}. These complex-valued instantons are called \textit{fuzzy instantons} \cite{Hartle2008a}. The fuzziness of instantons has interesting applications \cite{Chen:2015ria}. One example is the possibility of non-compact instanton solutions, referred to as fuzzy Euclidean wormholes \cite{Chen:2016ask,Chen:2018atc}.

Here we briefly explain the reason for such a solution. Let us first assume that the potential is flat, i.e., $V(\phi) = V_{0} > 0$, for simplicity. Then the scalar field equation, Eq.~(\ref{eq:2}), becomes
\begin{eqnarray}
\frac{\ddot{\phi}}{\dot{\phi}} = - 3 \frac{\dot{a}}{a},
\end{eqnarray}
which is exactly solvable:
\begin{eqnarray}
\frac{d\phi}{dt} = \frac{\mathcal{A}}{a^{3}}
\end{eqnarray}
with a constant $\mathcal{A}$. One important comment is that if we Wick-rotate in the manner that by $dt = -i d\tau$ and we require that $\dot{\phi}$ be real in Lorentzian signatures, then these conditions imply that $\dot{\phi}$ must be purely imaginary in Euclidean signatures, following the relation
\begin{eqnarray}
\frac{d\phi}{d\tau} = - i \frac{\mathcal{A}}{a^{3}}.
\end{eqnarray}

Therefore, the equation for $a$ in Euclidean signatures becomes
\begin{eqnarray}
\dot{a}^{2} + V_{\mathrm{eff}}(a) &=& 0,\\
V_{\mathrm{eff}}(a) &=& - 1 + \frac{8\pi}{3} \left( \frac{\mathcal{A}^{2}}{2a^{4}} + V_{0} a^{2} \right),
\end{eqnarray}
where $V_{\mathrm{eff}}(a) < 0$ is the physically allowed region. For simplicity, one can define  $a_{0}=(4\pi \mathcal{A}^{2}/3)^{1/4}$ and $\ell = (3/8\pi V_{0})^{1/2}$. Then, there can be two solutions of $V(a)=0$, say $a_{\mathrm{min}}$ and $a_{\mathrm{max}}$, and the physical solutions are allowed between $a_{\mathrm{min}} \leq a \leq a_{\mathrm{max}}$. In this case, the probability is \cite{Chen:2016ask,Chen:2018atc}
\begin{eqnarray}
\log P \simeq - S_{\mathrm{E}} \simeq 3 \pi \int_{a_{\mathrm{min}}}^{a_{\mathrm{max}}} da \frac{a \left(1 - \frac{a^{2}}{\ell^{2}} \right)}{\sqrt{1 - \left( \frac{a_{0}^{4}}{a^{4}} + \frac{a^{2}}{\ell^{2}} \right)}} \simeq \pi \ell^{2} \left[ 1 + 0.16 \left(\frac{a_{0}}{\ell}\right)^{5/2} \right] \label{eq:up},
\end{eqnarray}
where the action integral covers the full Euclidean time period. One can verify that for a given $\ell$, the probability of the Euclidean wormholes is much larger than that of compact instantons. In general, there exist much wider range in the parameter spaces for Euclidean wormhole solutions. We caution, however, that this is the maximum probability that a Euclidean wormhole can have \cite{Chen:2016ask}.

\begin{figure}[htb]
\centering
\includegraphics[width=0.6\textwidth]{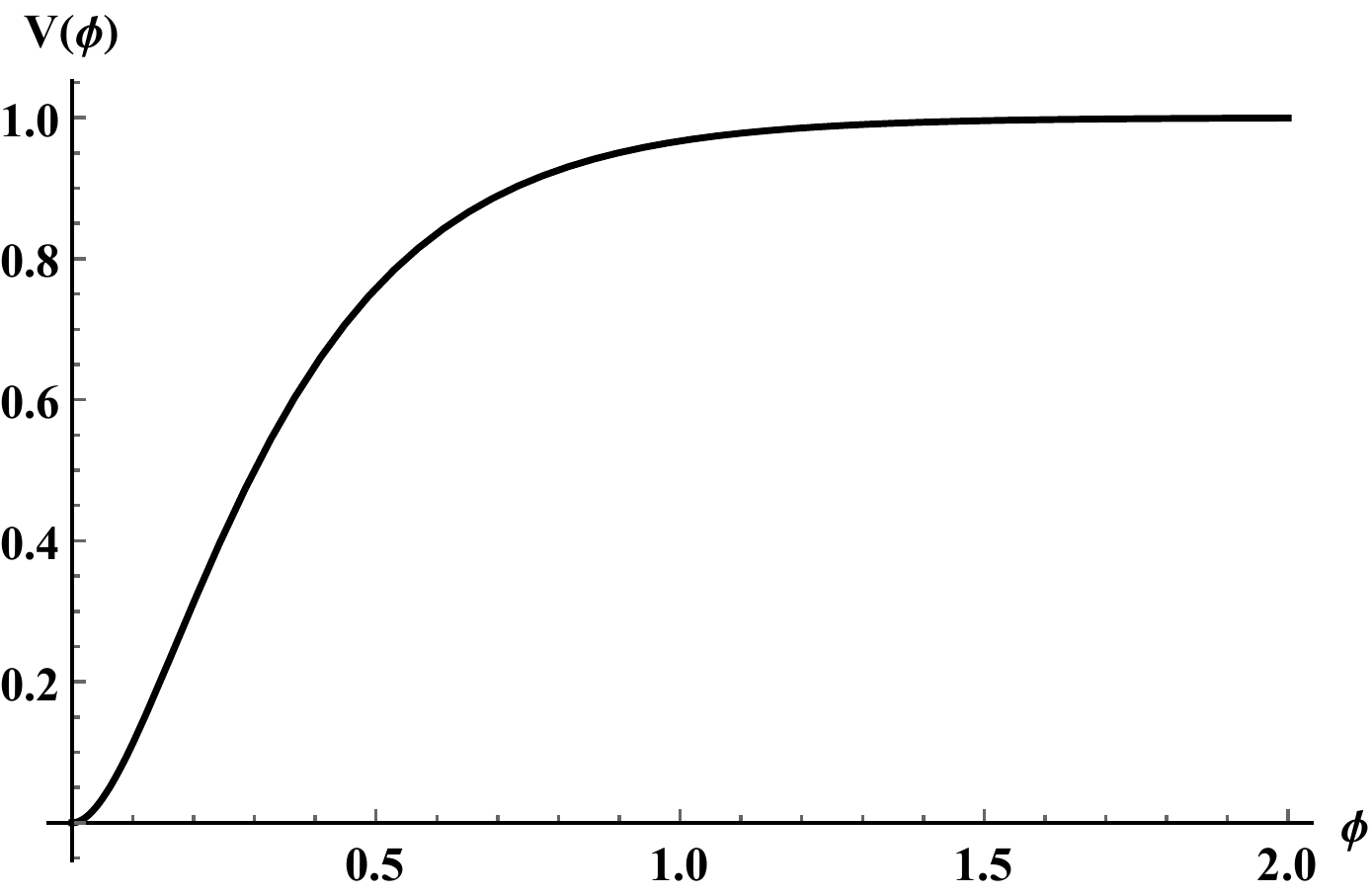}
\caption{The Starobinsky model.}
\label{fig:potential}
\end{figure}

\subsection{Classicality and necessity of a concave potential}

Not all the fuzzy instantons, however, are relevant to the creation of universes. After the Wick-rotation to the Lorentzian time, the manifold should be smoothly connected to the observer, who is supposed to be classical and only measures real-valued functions. Such a classical behavior is restored if the probability is slowly varies along the history \cite{Hartle2008a}.

Mathematically, one can present as follows. If we approximately write the wave function (using the steepest-descent approximation) as
\begin{equation}
\Psi[q_{I}] \simeq e^{- S_{\mathrm{re}}[q_{I}] + i S_{\mathrm{im}}[q_{I}]},
\end{equation}
where $q_{I}$ are canonical variables with $I=1,2,3, ...$, then the \textit{classicality condition} means that
\begin{equation} \label{eqn:classicality}
\left|\nabla_I S_{\mathrm{re}}\left[q_{I}\right]\right| \ll \left|\nabla_I S_{\mathrm{im}}\left[q_{I}\right]\right|,
\end{equation}
for all $I$. Then its history satisfies the semi-classical Hamilton-Jacobi equation \cite{Hartle2008a}. The intuitive meaning of the classicality condition is that when we solve on-shell Euclidean equations with complex-valued functions, such \textit{complex-valued functions should migrate to real values} after the Wick-rotation and a sufficient Lorentzian time, where detailed discussions are available in the appendix of \cite{Hwang2015}.

\begin{figure}[htb]
\centering
\subfigure[][]{\includegraphics[width=0.4\textwidth,height=0.18\textheight]{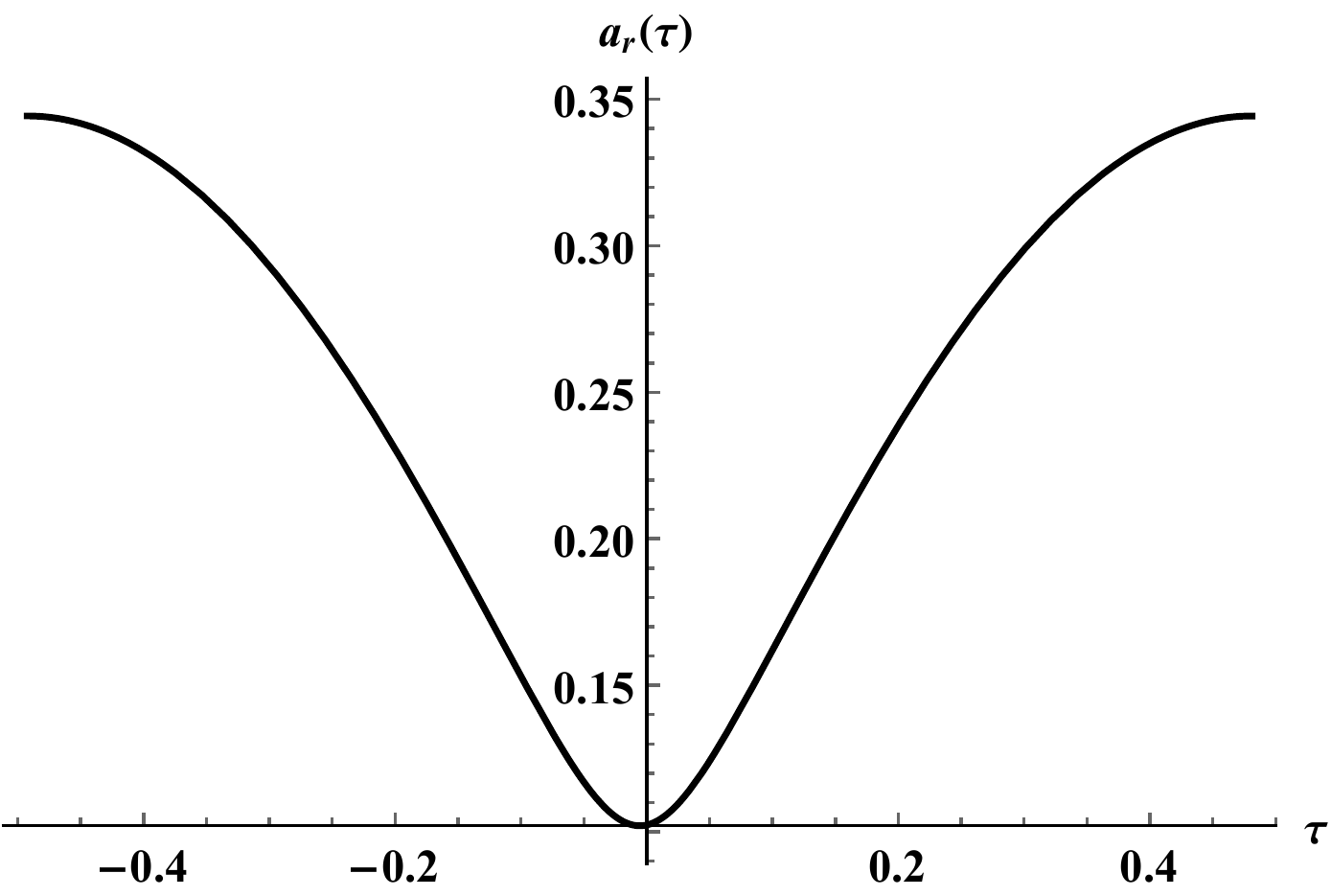}}
\subfigure[][]{\includegraphics[width=0.4\textwidth,height=0.18\textheight]{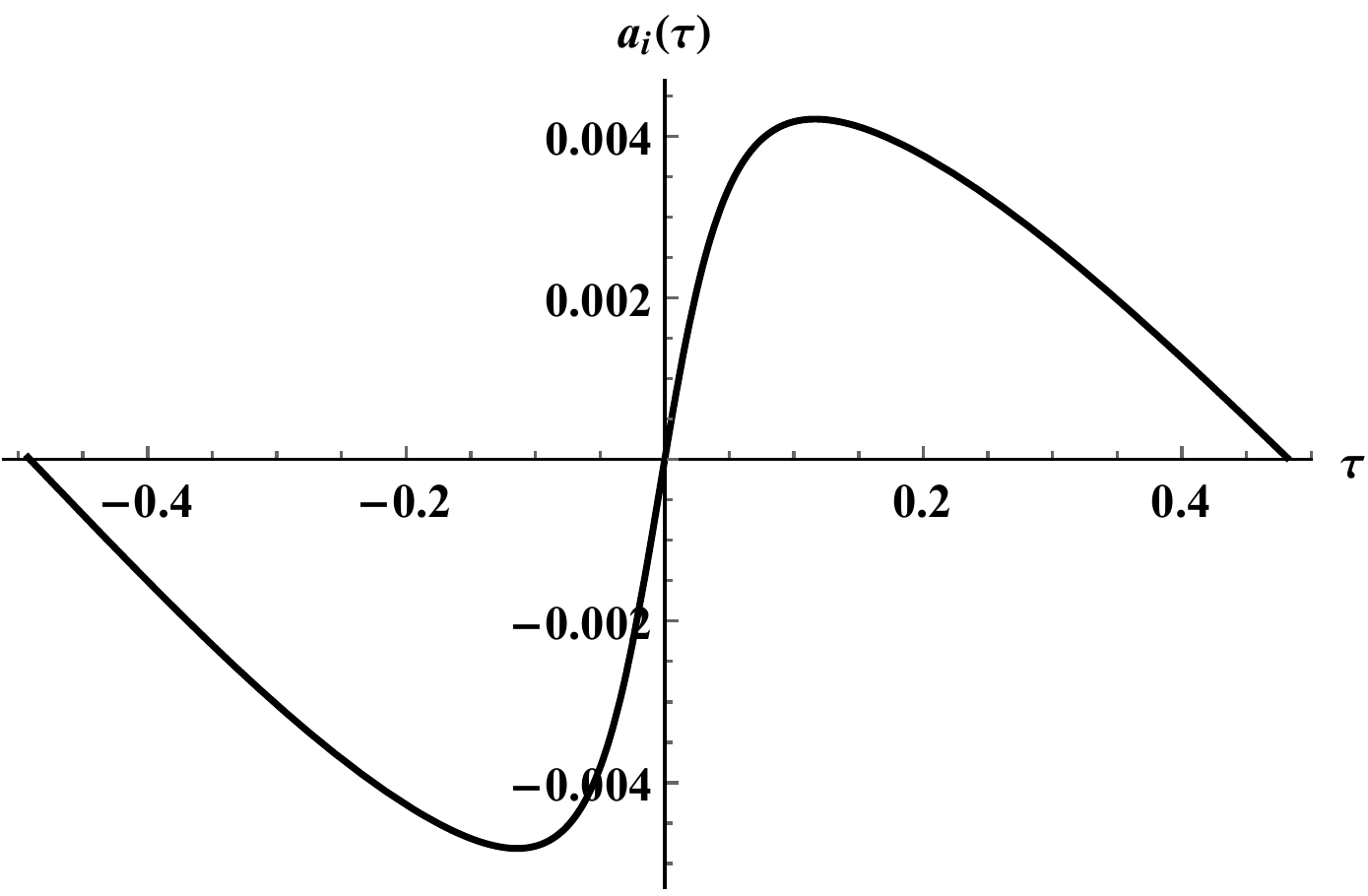}}
\subfigure[][]{\includegraphics[width=0.4\textwidth,height=0.18\textheight]{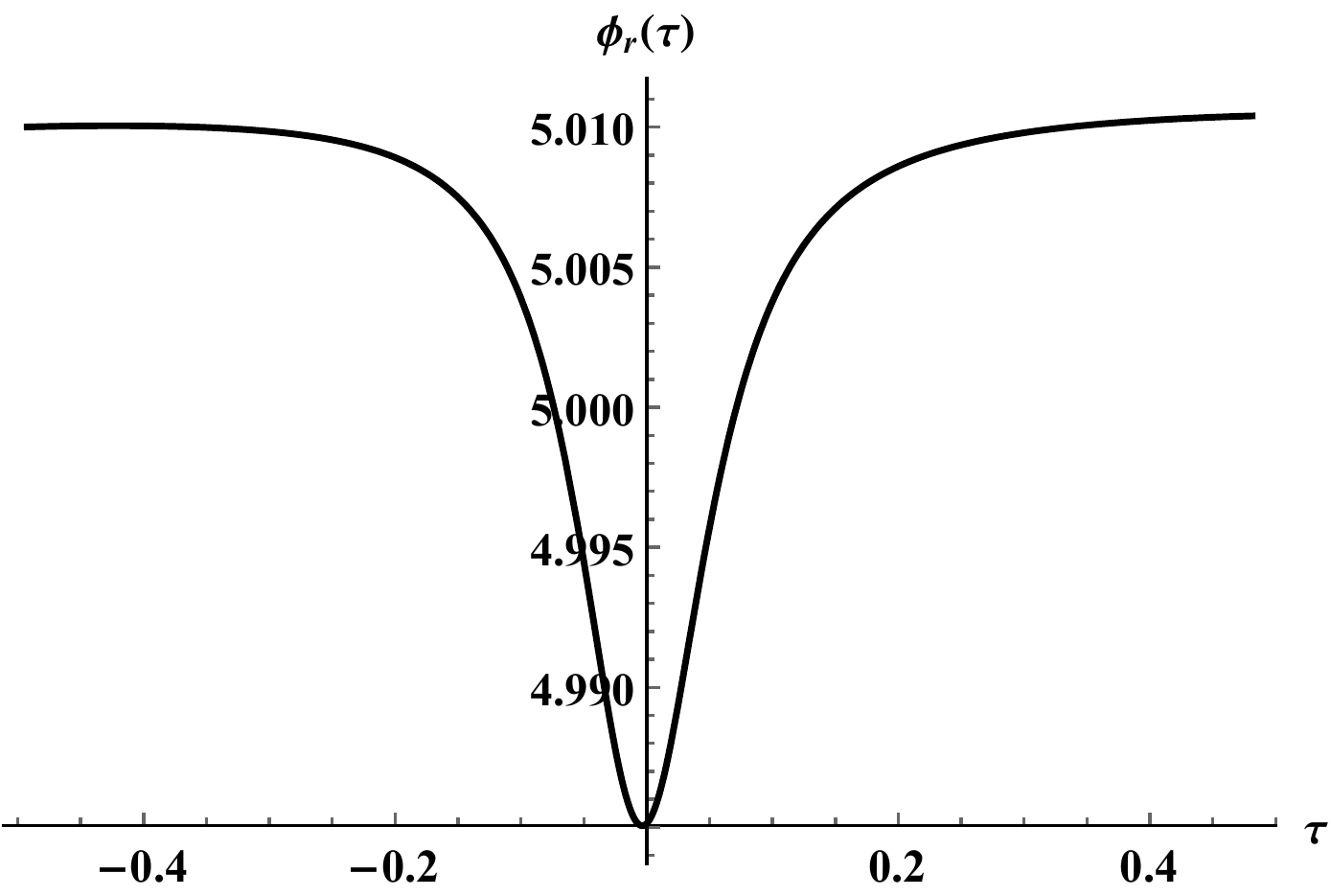}}
\subfigure[][]{\includegraphics[width=0.4\textwidth,height=0.18\textheight]{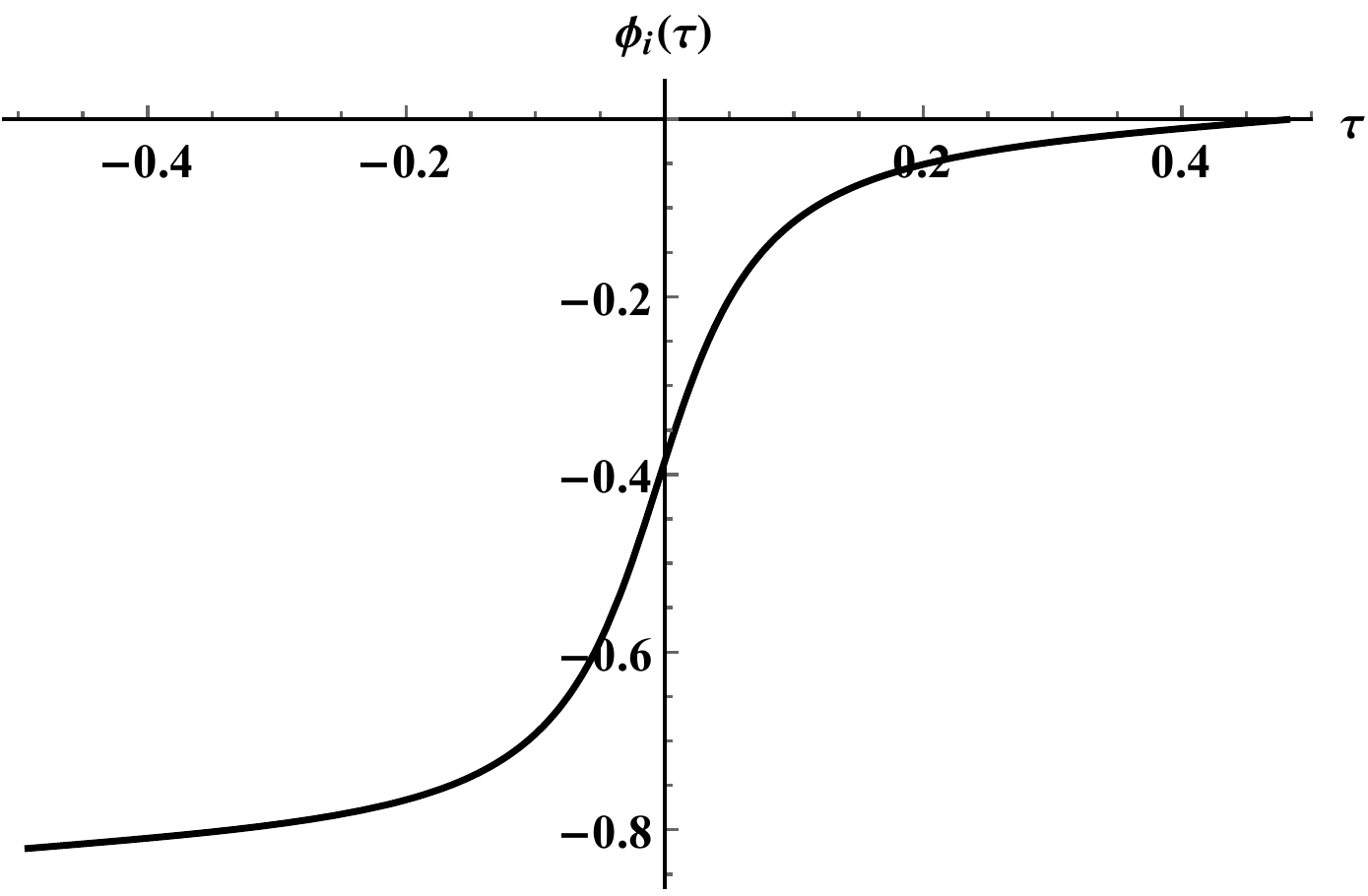}}
\caption{An example of the solution on the Euclidean domain.}
\label{fig:example_1_eu}
\end{figure}

\begin{figure}[p]
\centering
\subfigure[][]{\includegraphics[width=0.4\textwidth,height=0.18\textheight]{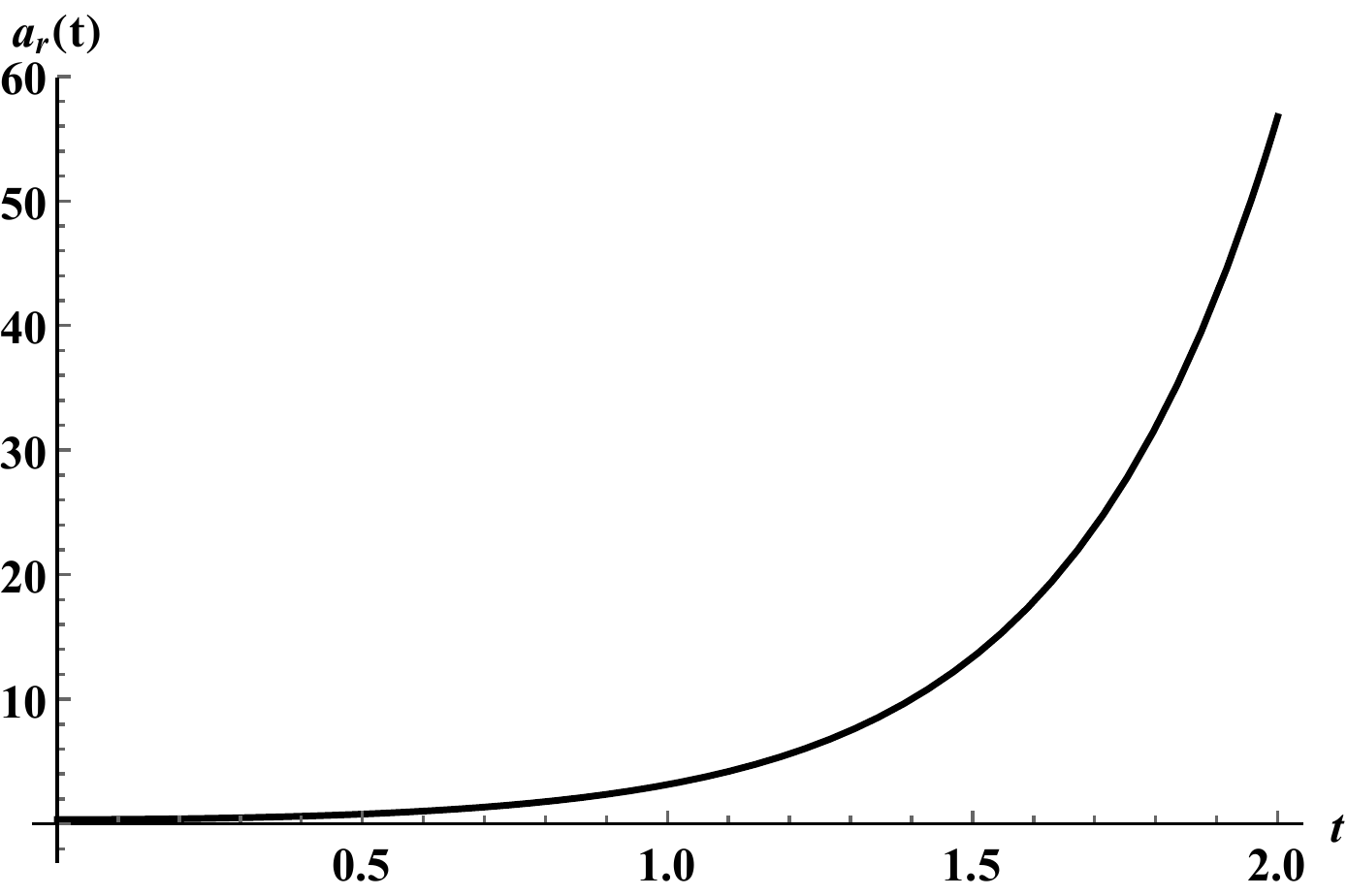}}
\subfigure[][]{\includegraphics[width=0.4\textwidth,height=0.18\textheight]{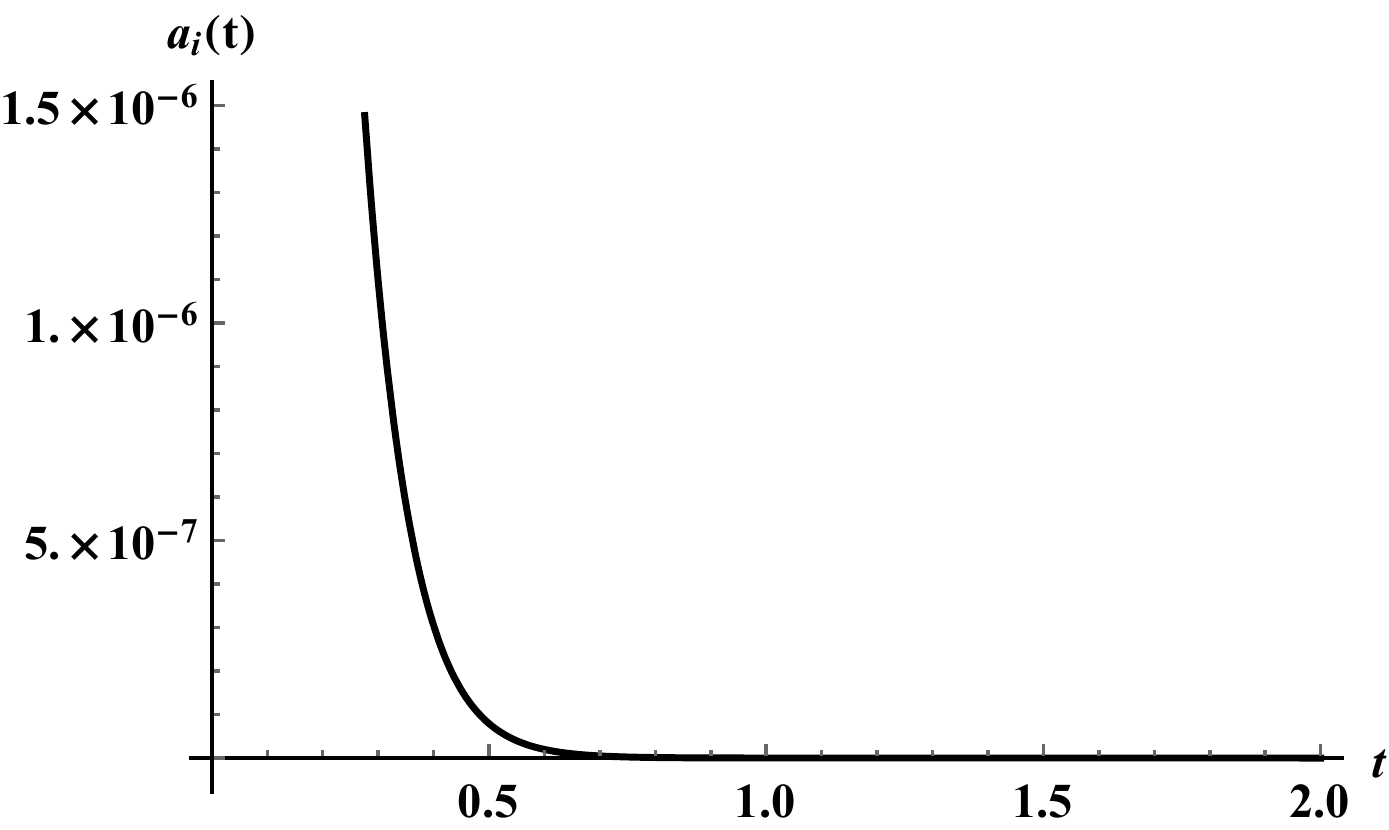}}
\subfigure[][]{\includegraphics[width=0.4\textwidth,height=0.18\textheight]{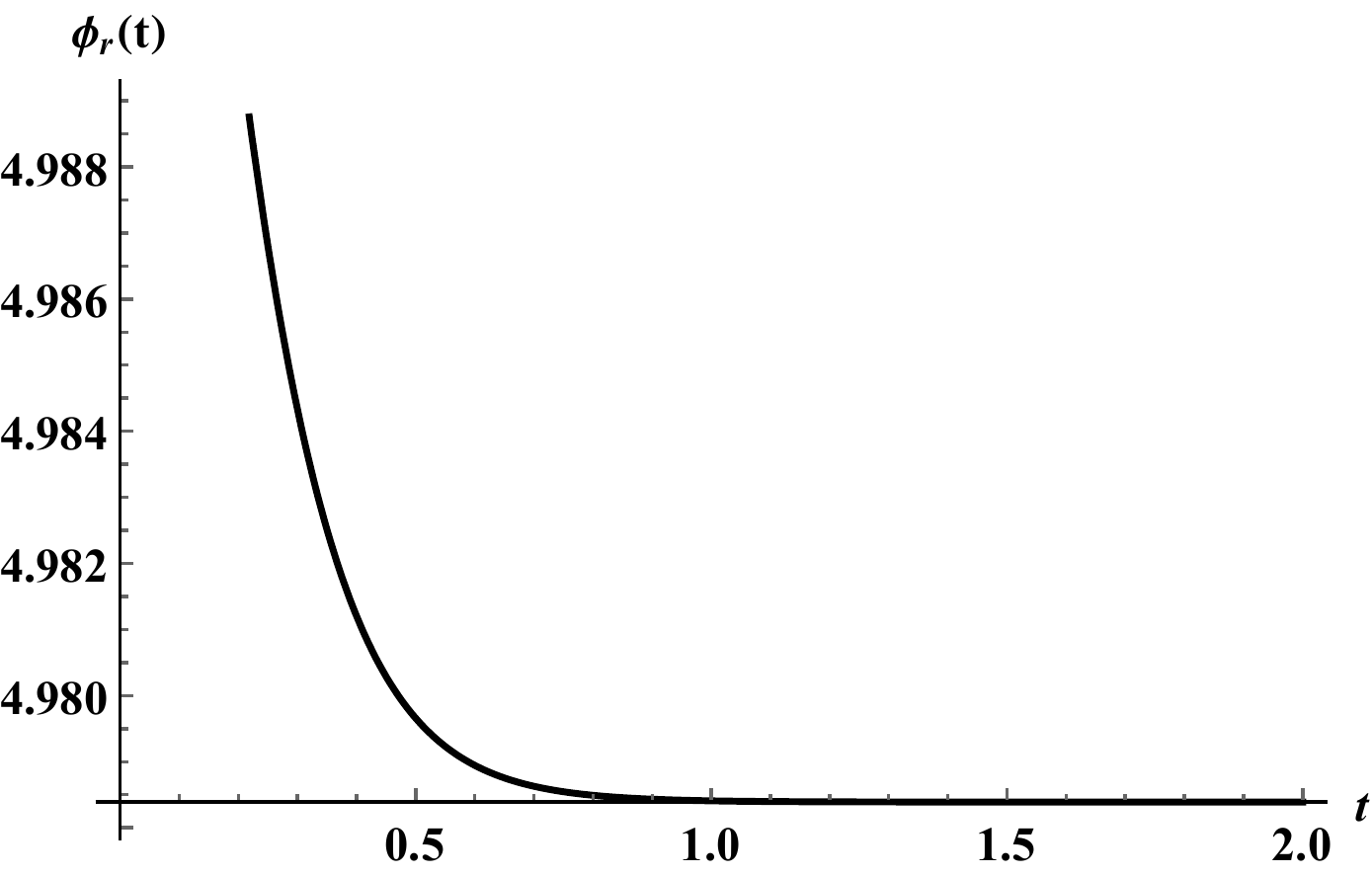}}
\subfigure[][]{\includegraphics[width=0.4\textwidth,height=0.18\textheight]{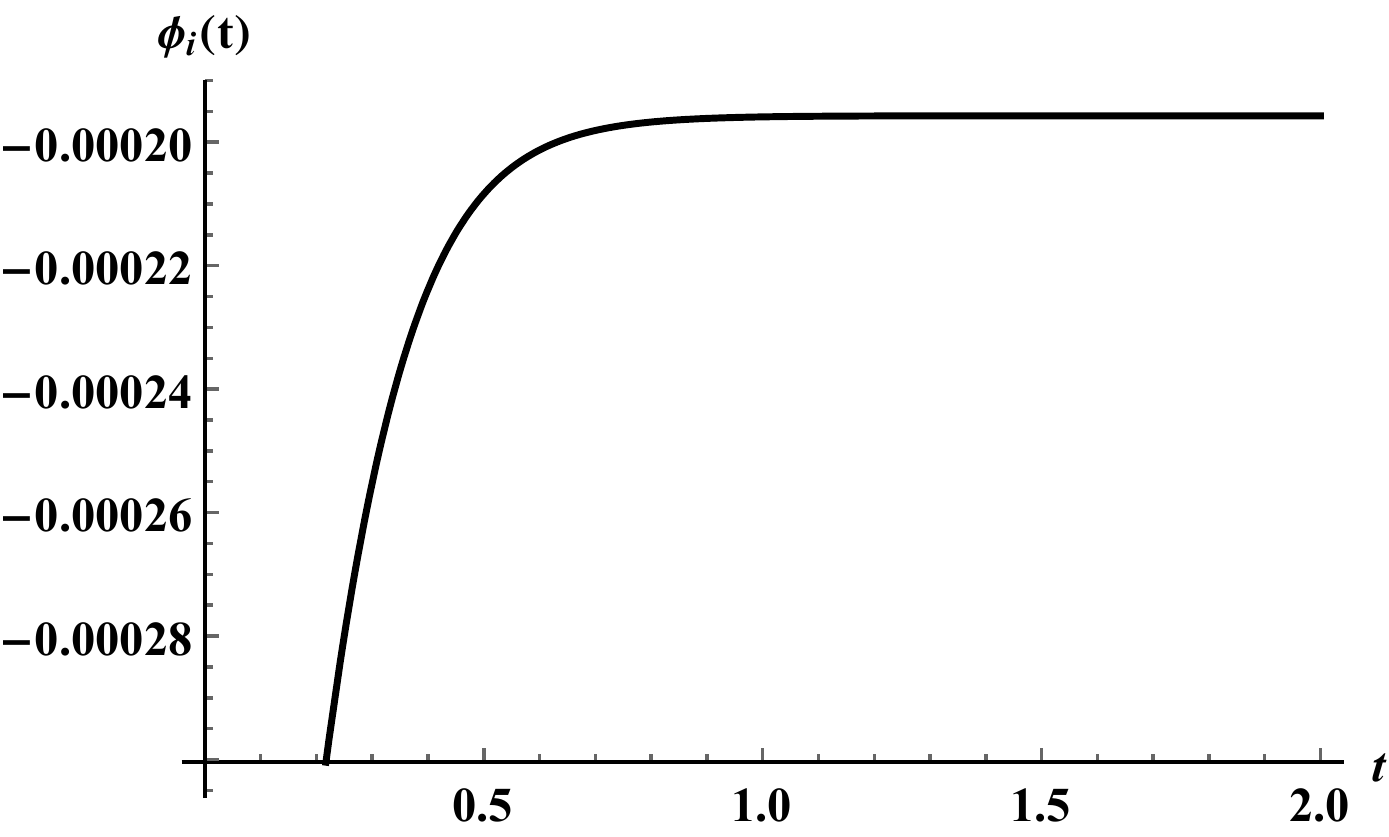}}
\caption{One end of the Euclidean wormhole after the Wick-rotation. This universe is classicalized since $a_{i}$ and $\phi_{i}$ approach to zero.}
\label{fig:example_1_lop}
\centering
\subfigure[][]{\includegraphics[width=0.4\textwidth,height=0.18\textheight]{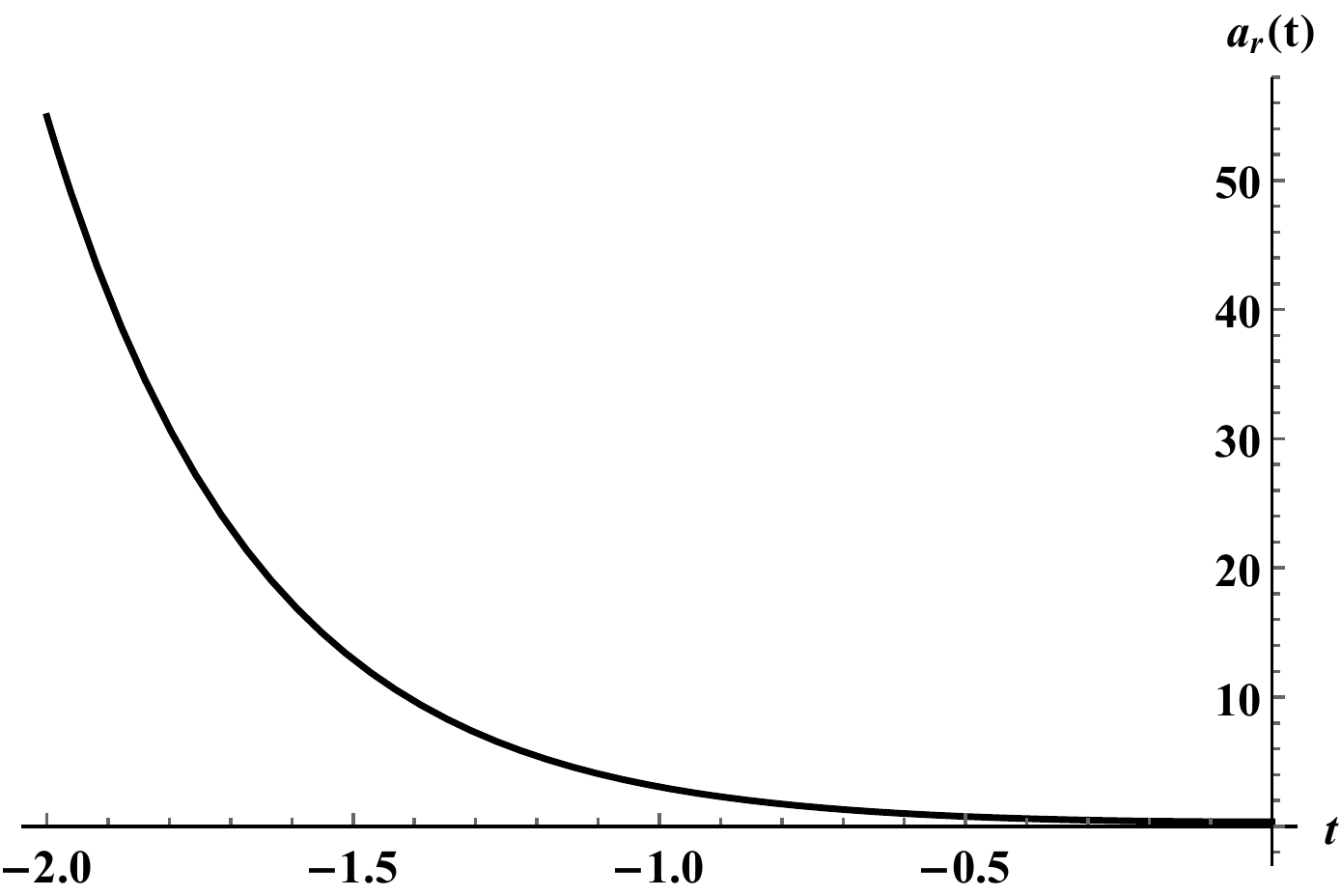}}
\subfigure[][]{\includegraphics[width=0.4\textwidth,height=0.18\textheight]{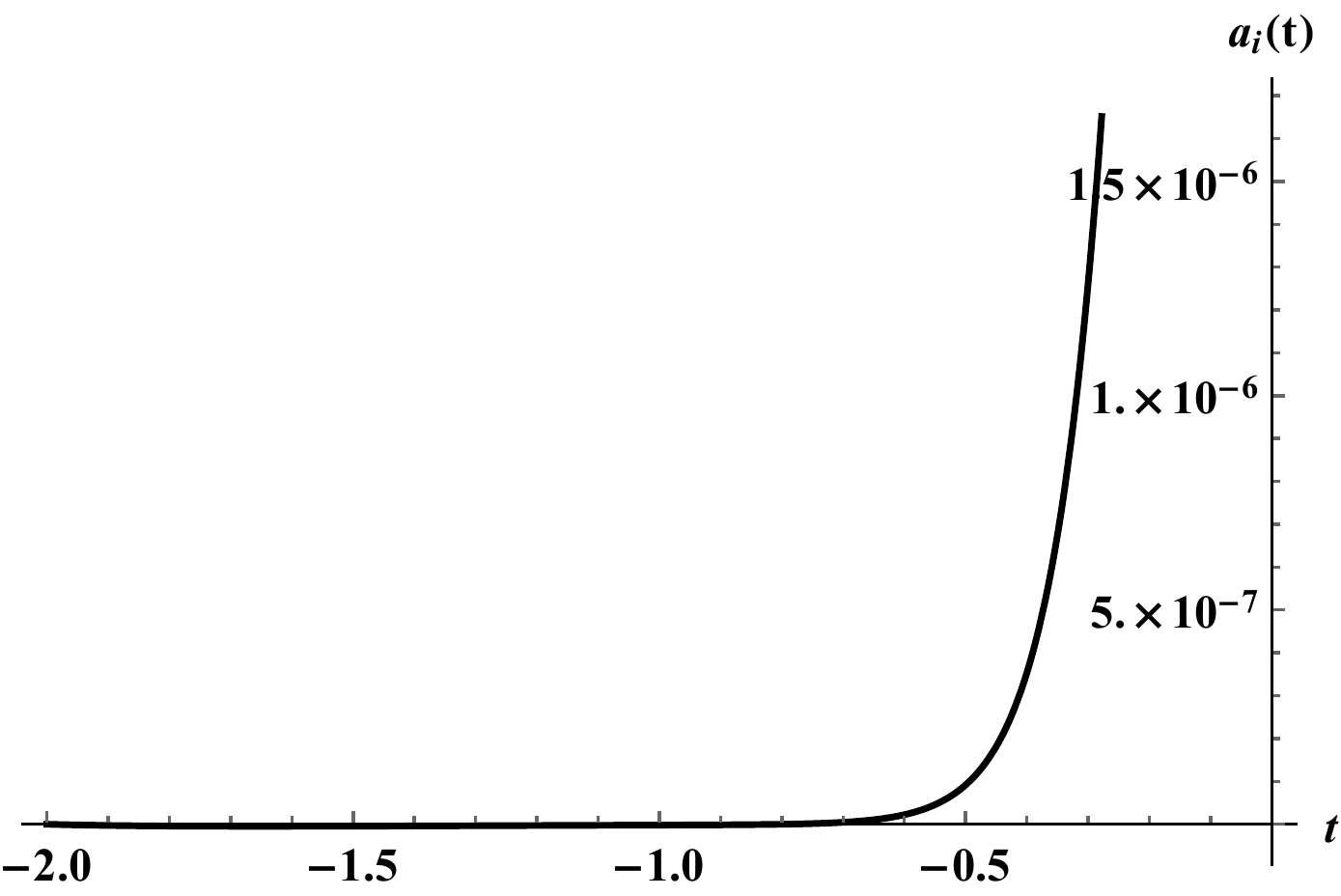}}
\subfigure[][]{\includegraphics[width=0.4\textwidth,height=0.18\textheight]{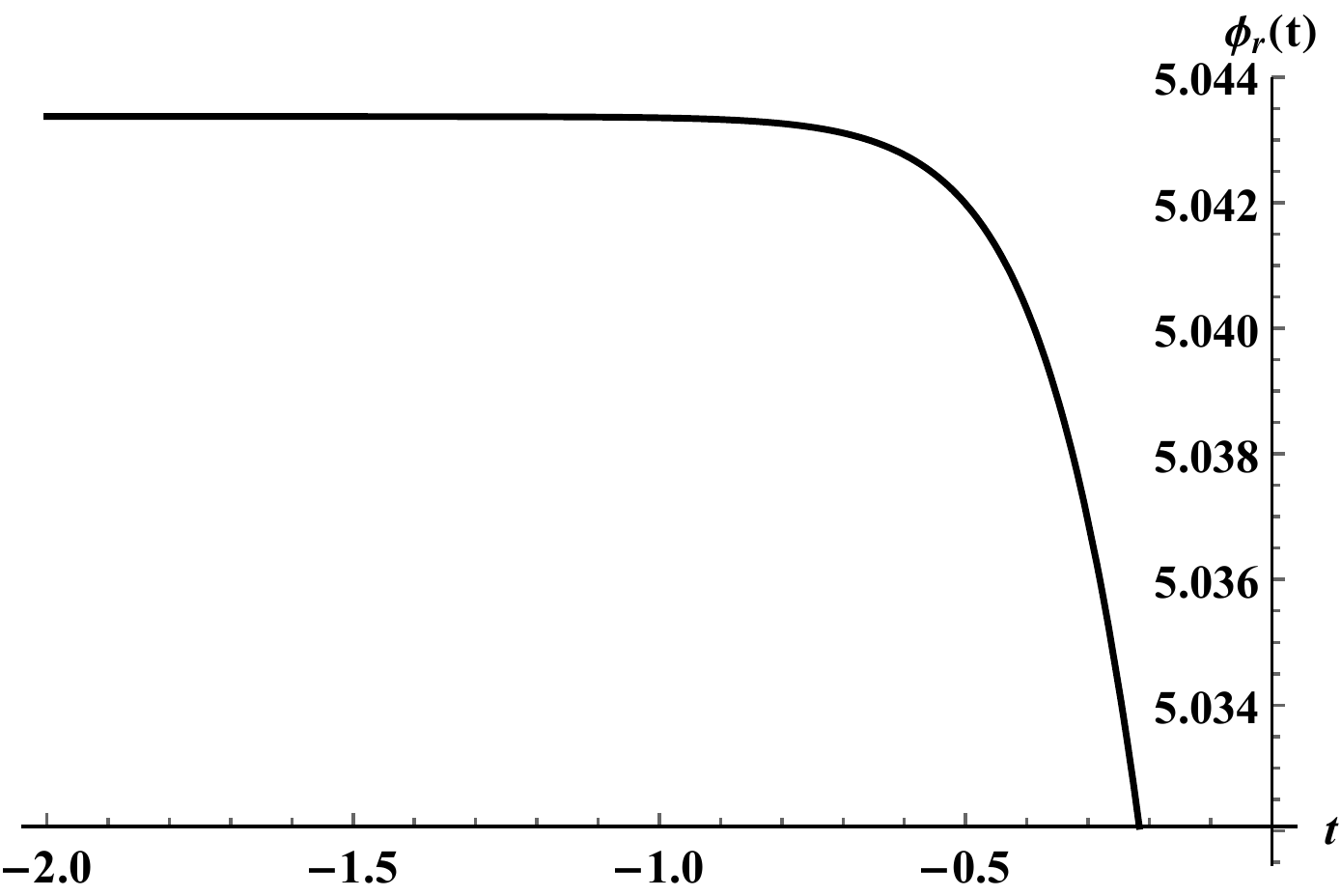}}
\subfigure[][]{\includegraphics[width=0.4\textwidth,height=0.18\textheight]{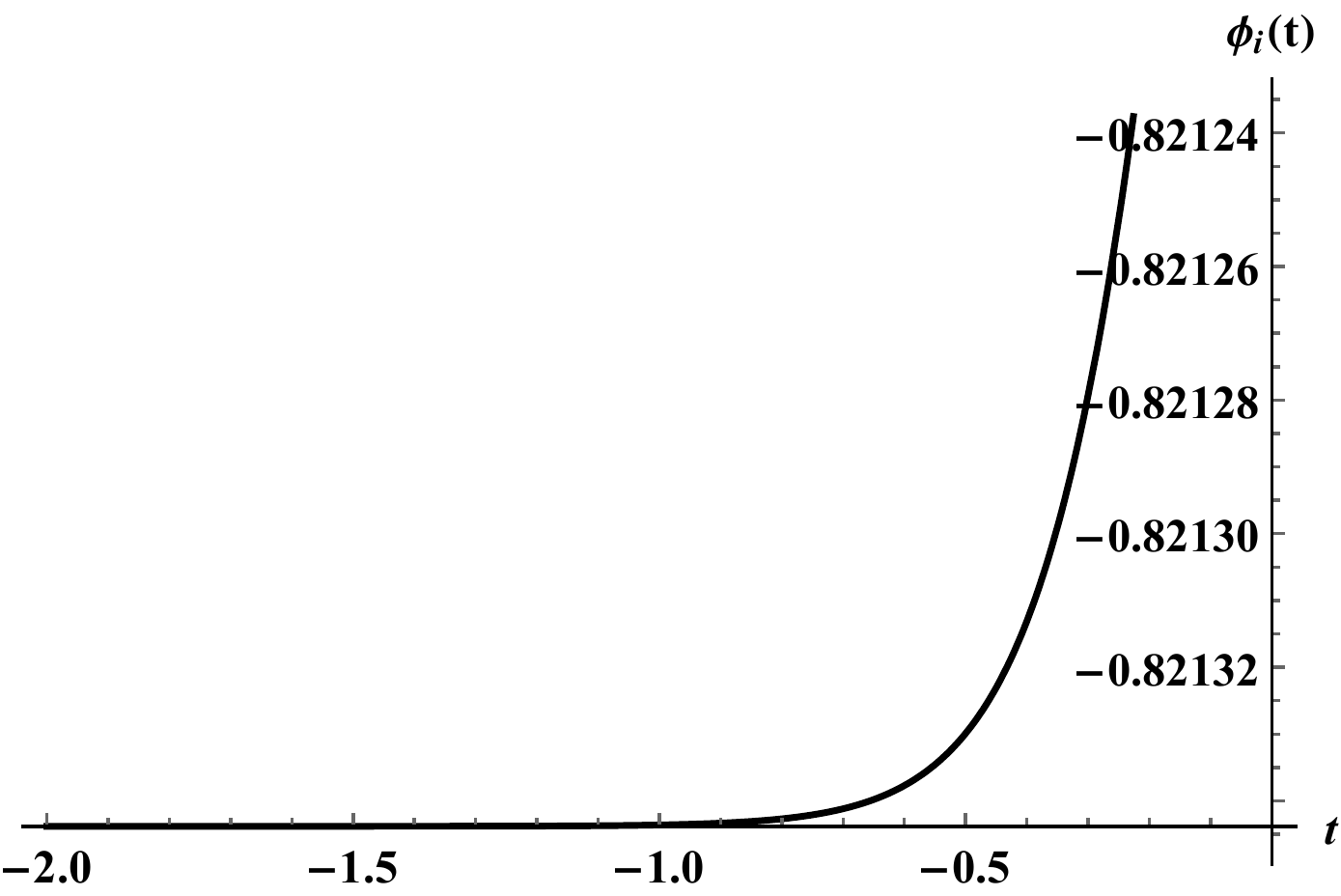}}
\caption{The other end of the Euclidean wormhole after the Wick-rotation. This universe is also classicalized since $a_{i} \rightarrow 0$ and $\phi_{i} \rightarrow \mathrm{constant}$.}
\label{fig:example_1_lon}
\end{figure}

For compact instantons, one only needs to impose the classicality condition at one end (final boundary). As long as the slow-roll condition is satisfied, it has been shown that this can be attained with very general inflation models \cite{Hwang2012}. However, for Euclidean wormholes, one needs to impose the classicality for both ends (initial and final boundaries) \cite{Chen:2018atc}. Of course, if the classicality condition for both ends are not satisfied, then the probability of Euclidean wormholes are not well-defined, and hence it has no sensible meaning in quantum cosmology. Such a demand translates into a constraint on a particular form of the inflaton potential. We briefly describe the reason as follows.

In order to give a consistent initial condition, we introduce the ansatz at $\tau = 0$. First, we impose non-vanishing $a_{0}$ at the bottleneck:
\begin{eqnarray}
&&a_{r}(0) = a_{\mathrm{min}} \cosh \eta,\\
&&a_{i}(0) = a_{\mathrm{min}} \sinh \eta,
\end{eqnarray}
$r$ and $i$ denote the real and the imaginary part, respectively, and $a_{\mathrm{min}}$ and $\eta$ are free parameters. This solution is possible due to the complexification of the scalar field:
\begin{eqnarray}
&&\phi_{r}(0) = \phi_{0} \cos \theta,\\
&&\phi_{i}(0) = \phi_{0} \sin \theta,\\
&&\dot{\phi}_{r}(0) = \frac{\mathcal{A}}{a_{\mathrm{min}}^{3}} \sinh{\zeta},\\
&&\dot{\phi}_{i}(0) = \frac{\mathcal{A}}{a_{\mathrm{min}}^{3}} \cosh{\zeta},
\end{eqnarray}
where $\phi_{0}$ is the modulus of the field amplitude, $\theta$ is the initial phase, and the time derivative of the field has two degrees of freedom due to two free parameters $\mathcal{A}$ and $\zeta$. In order to satisfy the constraint equation Eq.~(\ref{eq:1}), although it is not the most generic choice, it is convenient to choose the form for $\dot{a}$ as
\begin{eqnarray}
&&\dot{a}_{r}(0) = \sqrt{\frac{4\pi}{3}} \frac{\mathcal{A}}{a_{\mathrm{min}}^{2}} \sqrt{\sinh{\zeta}\cosh{\zeta}},\\
&&\dot{a}_{i}(0) = \sqrt{\frac{4\pi}{3}} \frac{\mathcal{A}}{a_{\mathrm{min}}^{2}} \sqrt{\sinh{\zeta}\cosh{\zeta}}.
\end{eqnarray}
The constraint equation, Eq.~(\ref{eq:1}), further restricts $a_{\mathrm{min}}$ and $\eta$:
\begin{eqnarray}
&& \label{eq:c1}0 = 1 + \frac{8\pi}{3} a_{\mathrm{min}}^{2} \left( -V_{\mathrm{re}} + 2 \cosh \eta \sinh \eta \; V_{\mathrm{im}} \right) - \frac{8\pi \mathcal{A}^{2}}{3 a_{\mathrm{min}}^{4}} \left( \frac{1}{2} + 2 \cosh \zeta \sinh \zeta \cosh \eta \sinh \eta \right),\\
&& \label{eq:c2}a_{\mathrm{min}}^{6} = \frac{\mathcal{A}^{2} \cosh \eta \sinh \eta}{-2 \cosh \eta \sinh \eta \; V_{\mathrm{re}} - V_{\mathrm{im}}},
\end{eqnarray}
where $V_{\mathrm{re}}$ and $V_{\mathrm{im}}$ denote the real and the imaginary part of the potential, respectively.

In order to satisfy the classicality condition, $\theta$ may be specified. Hence, for a given turning time $X$, through $\tau = X + it$, as $t \rightarrow \infty$, one may impose $a_{i} \rightarrow 0$ and $\phi_{i} \rightarrow 0$. However, the other parameters, $\mathcal{A}$, $\zeta$, and $\phi_{0}$, are nothing but just characterizing the shape of the wormhole; $\mathcal{A}$ is related to the size of the throat, $\zeta$ is used to tune the asymmetry of both sides of the wormhole, and $\phi_{0}$ corresponds to the initial field value on a given potential. Therefore, it is in general difficult to satisfy the classicality condition $a_{i} \rightarrow 0$ and $\phi_{i} \rightarrow 0$ at the other end.

The next alternative would be to slightly extend the notion of the classicality to require $\phi_{i} \rightarrow \mathrm{constant}$, under which $a_{i} \rightarrow 0$ follows automatically. Although the imaginary value of the field is non-zero, if the potential is sufficiently flat, then the observer of the second end of the wormhole will have the shift-symmetry of the field, and hence, after redefining the field value, the imaginary field value itself will be negligible after the Wick-rotation \cite{Chen:2016ask}; the observable degrees of freedom are $\dot{\phi}$, $a$, and $\dot{a}$, which are guaranteed to be sufficiently real. This is possible if the potential has a flat direction. This indicates that the classicalized fuzzy Euclidean wormholes can exist only on concave potentials.

\begin{figure}[htb]
\centering
\includegraphics[scale=0.75]{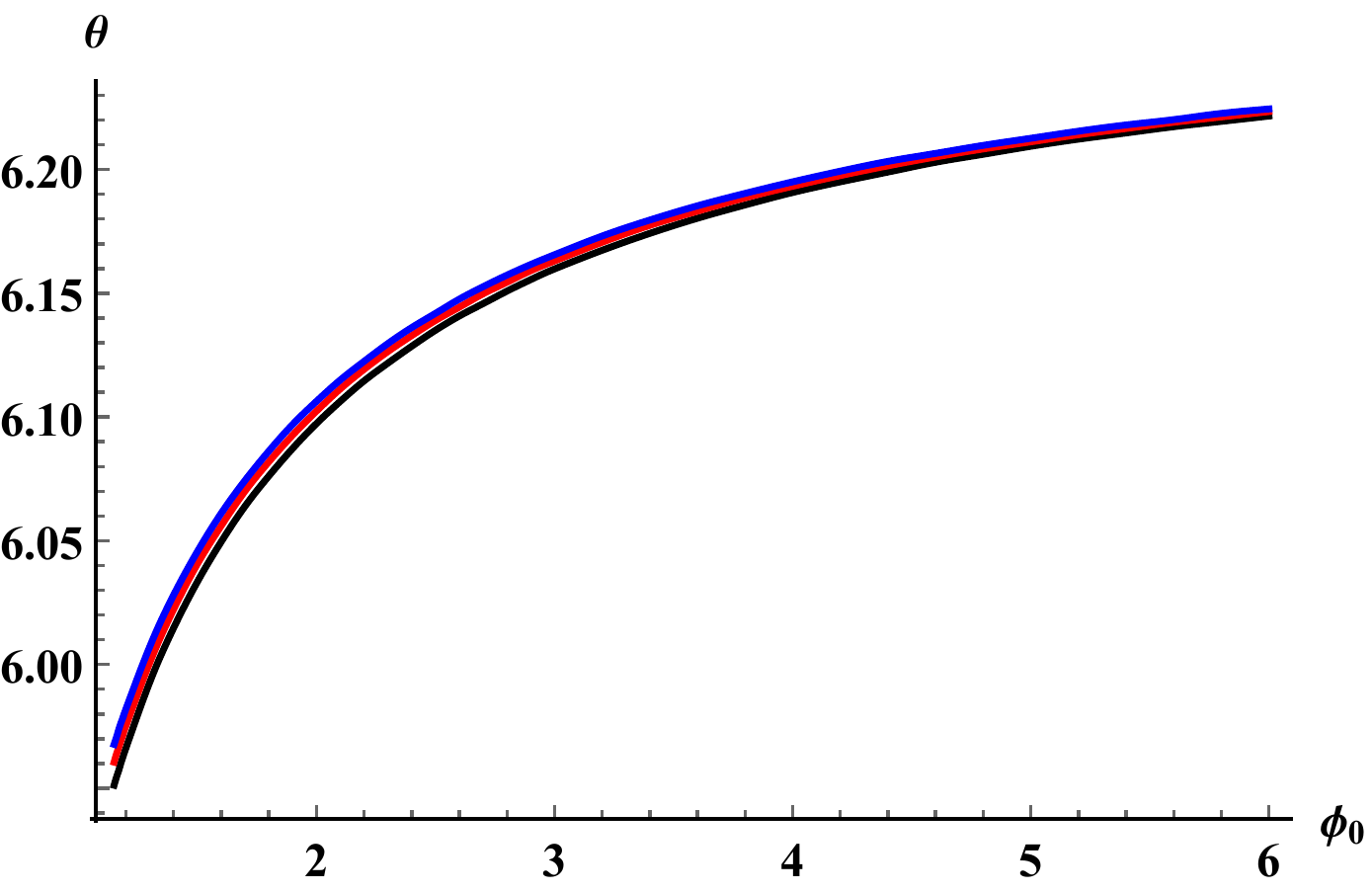}
\caption{$\theta$ vs. $\phi_{0}$, where $a_{0} = 0.05$ (black), $0.1$ (red), and $0.15$ (blue) with $\zeta = 0.01$.}
\label{fig:theta}
\centering
\includegraphics[scale=0.75]{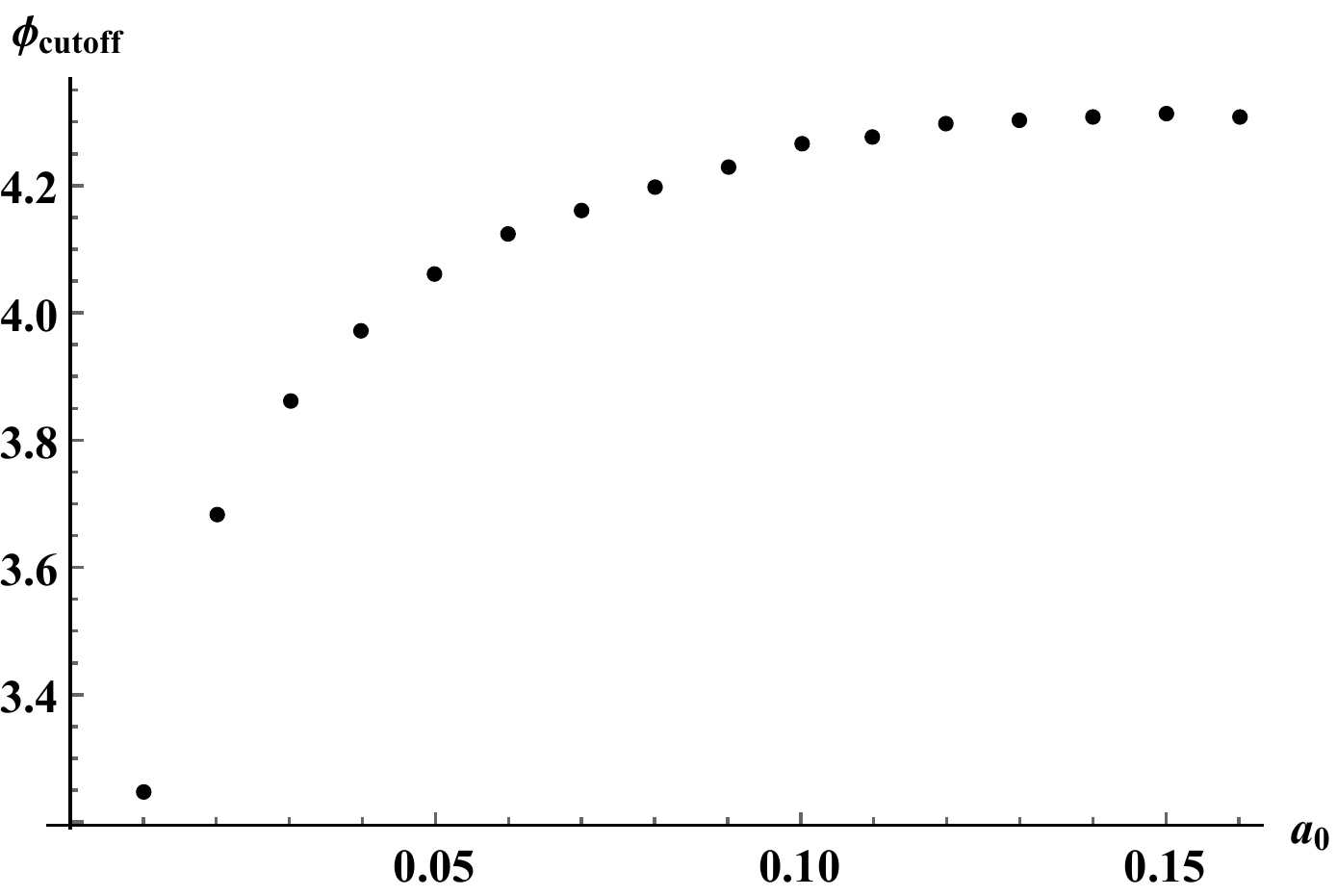}
\caption{$\phi_{\rm{cutoff}}$ by varying $a_{0}$.}
\label{fig:cutoff}
\end{figure}

Interestingly, such a concave potential is consistent with the current cosmological observations. Let us focus on the Starobinsky model (Fig.~\ref{fig:potential}), which successfully explains the CMB observations:
\begin{equation} \label{eq:aperp} 
	V(\phi) = V_0 \left(1-e^{-\sqrt{\frac{16\pi}{3}}\phi}\right)^2.
\end{equation}
We see that the classicality at both ends of the wormhole is satisfied: see Figs.~\ref{fig:example_1_eu}, \ref{fig:example_1_lop}, and \ref{fig:example_1_lon} as an explicit example. As we vary the initial condition $\phi_{0}$, we see that $\theta$ also vary in order to satisfy the classicality (Fig.~\ref{fig:theta}).

\section{\label{sec:Euc}Euclidean wormholes with the Starobinsky model}

Now let us discuss the probability distribution of fuzzy Euclidean wormholes with a specific inflaton potential. In order to do this, we have to vary the initial conditions and show the probability as a function of $\phi_{0}$ for a given set of ($a_{0} =(4\pi \mathcal{A}^{2}/3)^{1/4}$, $\zeta$). In addition, there will be a range of $\phi_{0}$ that can satisfy the classicality. For example, if the field value is too close to the local minimum, then both ends of the wormhole cannot be classicalized. We call such a bound as a cutoff $\phi_{\mathrm{cutoff}}$. By varying $a_{0}$ and $\zeta$, we will see the dependence of the cutoff.

\begin{figure}[htb]
\centering
\includegraphics[scale=0.55]{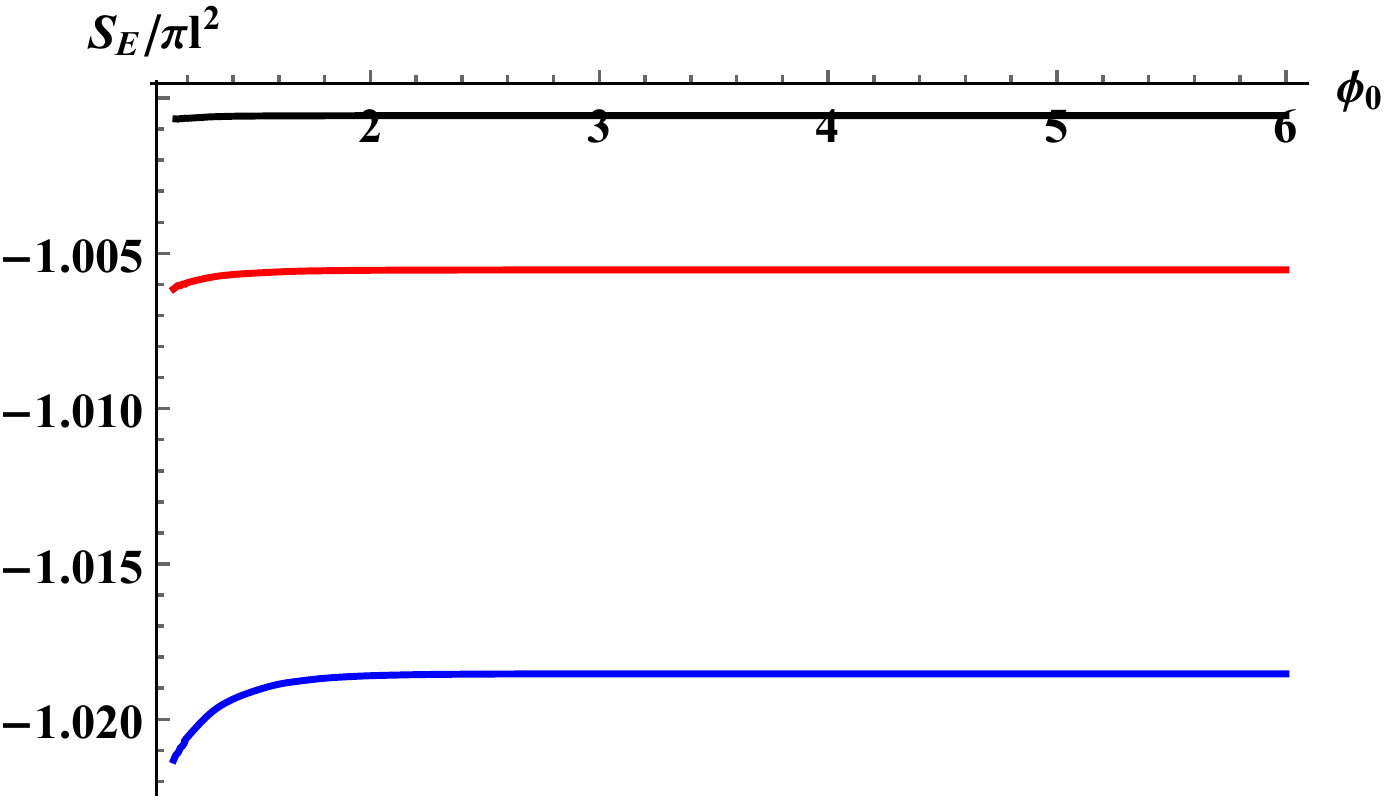}
\includegraphics[scale=0.55]{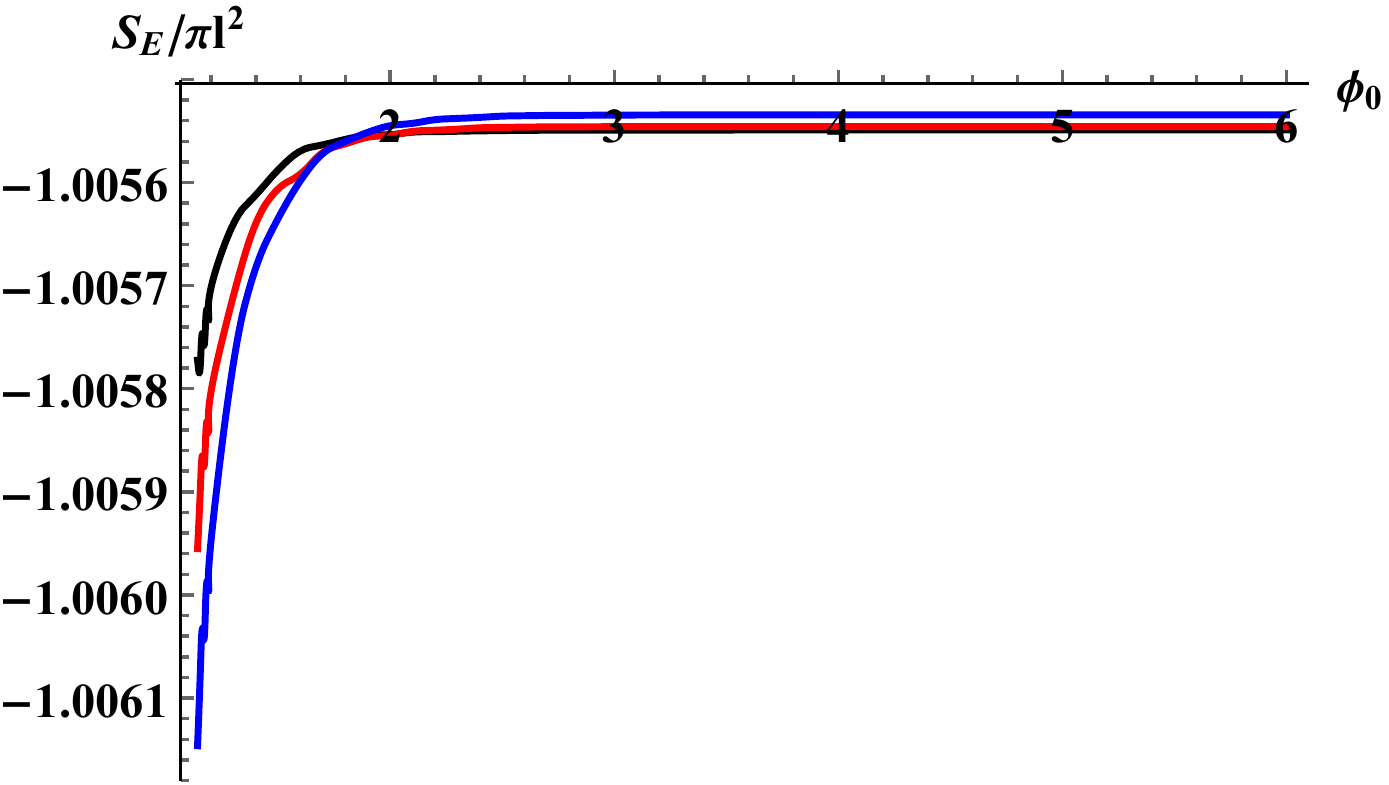}
\caption{Left: Euclidean action $S_{\mathrm{E}}/\pi\ell^{2}$ by varying $a_{0} = 0.05$ (black), $0.1$ (red), and $0.15$ (blue) fixing $\zeta = 0.01$. Right: Euclidean action $S_{\mathrm{E}}/\pi\ell^{2}$ by varying $\zeta = 0.1$ (black), $0.05$ (red), and $0.01$ (blue), by fixing $a_{0} = 0.1$. Note that the cutoff is $\phi \gtrsim 3$ and hence $\phi < 3$ has no deeper physical meaning.}
\label{fig:euaction}
\centering
\includegraphics[scale=0.75]{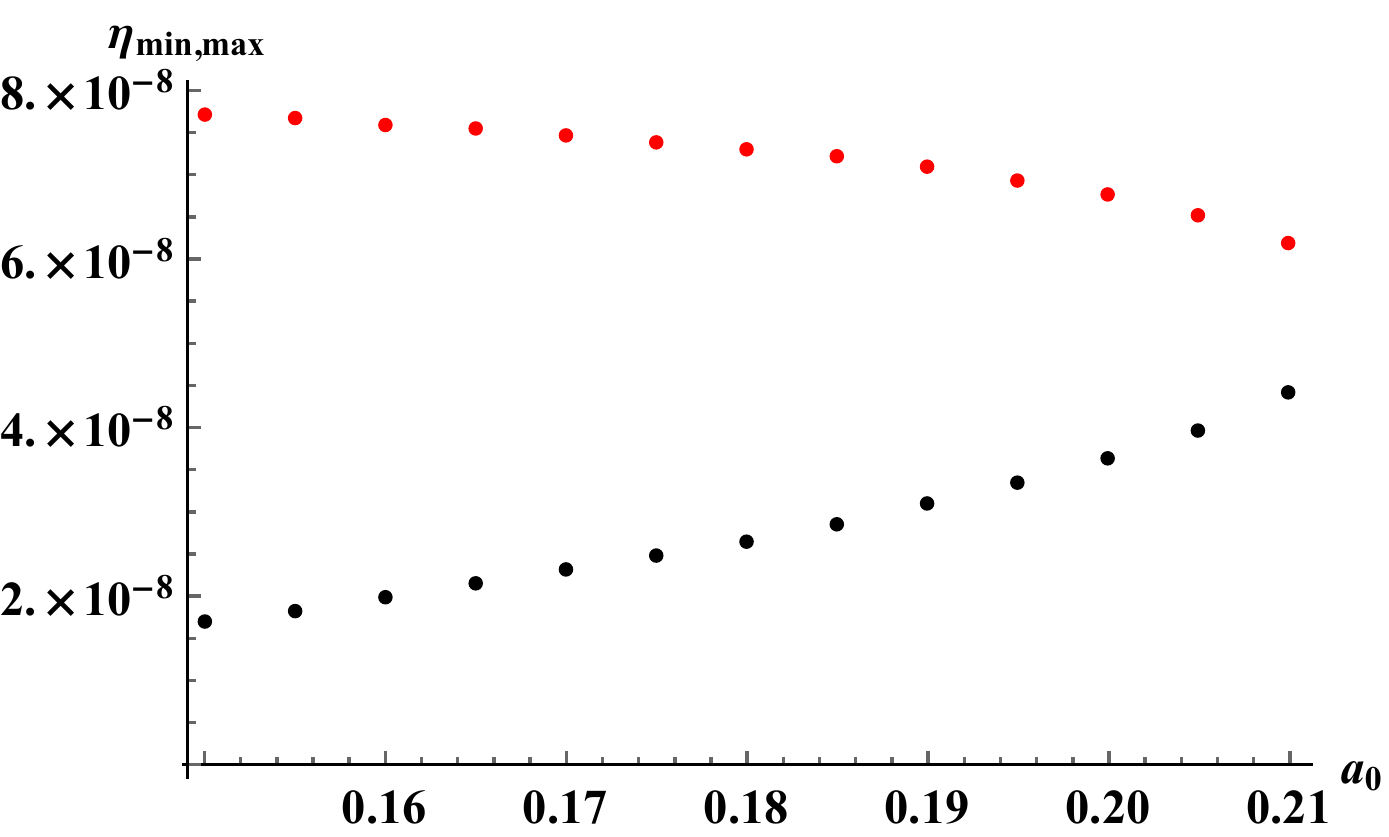}
\caption{$\eta_{\mathrm{min}}$ (black) and $\eta_{\mathrm{max}}$ (red) vs. $a_{0}$ by fixing $\phi_{0} = 4$ and $\zeta = 0.01$.}
\label{fig:eta}
\end{figure}

\subsection{Cutoffs and probabilities}

By varying initial conditions, we obtain a family of fuzzy Euclidean wormholes. At one end, the field and metric are classicalized by tuning the angle parameter $\theta$, while at the other end the scalar field is classicalized when it stops at the top of the hill.

Several comments are now in order. First, even though the scalar field stops at the hilltop of the potential, as time goes on, the field will receive thermal fluctuations and eventually role down into the local minimum. Hence, the numerical computation of too long Lorentzian time is not meaningful. We will only numerically compute by several orders of the Hubble time. Once the field is classicalized, the main contribution of the probability is well restricted to the Euclidean time and hence we integrate the action over the Euclidean section. The Starobinsky model has a single free parameter $V_{0} = 3 / 8\pi \ell^{2}$, but by rescaling the metric, i.e.,
\begin{equation}
ds_{\mathrm{E}}^{2} = \frac{1}{V_{0}} \left( d\tau^{2} + a^{2} d\Omega_{3}^{2} \right),
\end{equation}
one can make all the equations of motion independent of $\ell$. In terms of the dynamics of instantons, without losing generality, one can choose $V_{0} = 1$. The probability should include the $V_{0}$-dependence, and hence by presenting the action integral as $S_{\mathrm{E}}/\pi\ell^{2}$, one can show $V_{0}$-independent results.

For example, as we vary $a_{0}$, we can trace the change of cutoff $\phi_{\mathrm{cutoff}}$ in Fig.~\ref{fig:cutoff}. Note that whatever the other model parameters are, the cutoff is $\phi_{0} \sim \mathcal{O}(1)$. This means that the wormholes will be formed only the hilltop of the inflaton potential. We will see its physical meaning in the next subsection.

Fig.~\ref{fig:euaction} denotes the Euclidean action integration over the Euclidean domain and as we expected, as the initial field value decreases, the probability increases. Also, one can see that the value of $S_{\mathrm{E}}/\pi\ell^{2}$ is less than minus one; this implies that these wormholes are preferred than those of compact instantons. However, these solutions are only about the Euclidean domain; unless we impose the classicality after the Wick-rotation, these solutions do not have physical meaning.

It is easy to figure out that the large $a_{0}$ and small $\zeta$ limit are probabilistically preferred if $\phi > \phi_{\mathrm{cutoff}}$. The large $a_{0}$ limit corresponds that there exists a large contribution from the fuzzy scalar field. The small $\zeta$ limit corresponds that the wormhole becomes more and more symmetric \cite{Chen:2016ask}. So, these two tendencies do make sense. Note that in general the solutions of constraint equations Eqs.~(\ref{eq:c1}) and (\ref{eq:c2}) have two solutions of $\eta$, say $\eta_{\mathrm{min}}$ and $\eta_{\mathrm{max}}$, and we will choose the smaller one $\eta_{\mathrm{min}}$ as the initial condition ($\eta_{\mathrm{max}}$ corresponds the other turning point which corresponds the maximum radius). If $a_{0}$ increases too much, then $\eta_{\mathrm{min}}$ and $\eta_{\mathrm{max}}$ will disappear (Fig.~\ref{fig:eta}). So, there must be a bound of the probability as well as $a_{0}$.

\subsection{No-boundary vs. Euclidean wormholes}

It was shown that the probability of compact instantons (hypothesis $\mathcal{H}_{c}$) is approximately
\begin{eqnarray}
\log P\left[\phi_{0} | \mathcal{H}_{c} \right] \simeq \frac{3}{8 V(\phi_{0})},
\end{eqnarray}
where the most probable $\phi_{0}$ is near the cutoff $\phi_{0} = \phi_{c} \simeq 0.6$ \cite{Hwang:2012bd}. This generates only 2 or 3 $e$-foldings. On the other hand, the classicalized wormholes will appear only for $\phi_{0} \geq \phi_{\mathrm{cutoff}} \simeq \mathcal{O}(1)$ and hence the probability of the Euclidean wormholes is \cite{Chen:2018atc} (hypothesis $\mathcal{H}_{w}$)
\begin{eqnarray}
\log P \left[ \phi_{0} | \mathcal{H}_{w} \right] \simeq \pi \ell^{2} \left[ 1 + 0.16 \left(\frac{a_{0}}{\ell}\right)^{5/2} \right].
\end{eqnarray}
Note that $\log P\left[\phi_{0} | \mathcal{H}_{w} \right] \simeq 3/8V_{0}$ and $V(\phi_{c}) < V_{0}$ in general. Therefore, if we naively compare two instantons, then compact instantons with a very small number of $e$-foldings are exponentially preferred.

Note that this naive consideration of compact instantons is inconsistent with the CMB observations, which indicates that the inflation lasted for more than $50$ $e$-foldings. One way to ameliorate this difficulty is to insist that the number of $e$-foldings is larger than $50$, which translates into the constraint that $\phi_{0} > 1.1$ such that $V(\phi_{0} > 1.1) \sim V_{0}$:
\begin{eqnarray}
\log P\left[\phi_{0} | \mathcal{H}_{c}, \mathcal{H}_{\mathcal{N}} \right] \simeq \left.\frac{3}{8 V(\phi_{0})}\right|_{\phi_{0} > 1.1},
\end{eqnarray}
where now we introduced one more assumption about $e$-foldings ($\mathcal{H}_{\mathcal{N}}$).
As a result,
\begin{eqnarray}
\log P\left[ \phi_{0} | \mathcal{H}_{c}, \mathcal{H}_{\mathcal{N}} \right] \simeq \pi \ell^{2}.
\end{eqnarray}
In contrast, in the wormhole case $\phi_{\mathrm{cutoff}} > 1.1$ already, and so there is no need to impose such an ad hoc condition. Comparing these two probabilities, we see that
\begin{eqnarray}
\log \frac{P\left[ \phi_{0} | \mathcal{H}_{w}, \mathcal{H}_{\mathcal{N}} \right]}{P\left[ \phi_{0} | \mathcal{H}_{c}, \mathcal{H}_{\mathcal{N}} \right]} \simeq 0.16 \pi \ell^{2} \left(\frac{a_{0}}{\ell}\right)^{5/2} > 0.
\end{eqnarray}
In conclusion, since the inflation has undergone more than 50 $e$-foldings, Euclidean wormholes are more natural than the compact instantons to interpret the origin of the inflation.

\section{\label{sec:con}Conclusion}

In this paper, we investigated the Euclidean path integral with the Starobinsky inflation model. The Euclidean path integral is well approximated by instantons. We focused on two types of instantons. One is compact instantons and the other is non-compact instantons, the so-called Euclidean wormholes. For a given initial condition of the inflaton field, compact instantons are well defined unless it has several $e$-foldings. In comparison Euclidean wormhole solutions require boundary conditions at both ends. This in turn requires, as a necessary condition, the existence of a hilltop in the inflaton potential and the inflaton field should begin at a location near the hilltop.

A question naturally arises: which one is preferred over the other? At first glimpse, the compact instantons with a very small number of $e$-foldings are exponentially preferred. However, we already know that CMB data indicates that the inflation has undergone for more than $50$ $e$-foldings. This contradiction requires an explanation within the Euclidean path integral approach to quantum cosmology. In this paper we argue that such an explanation is indeed possible. We showed that once the constraint of $50$ $e$-foldings is imposed, the dominant instanton contributions are from Euclidean wormholes and not the compact ones. Note that such non-compact instanton solutions do not exist in the convex inflaton potential. This may explain why inflation models with concave potentials appear to be favored over those with convex potential in the Planck CMB data. 

It is interesting and important to explore the possible physical observables that can confirm or falsify the Euclidean wormhole hypothesis. Note, for example, that the perturbations of the manifold can be traced and their expectation values estimated. These values can then be checked against that generated from the Bunch-Davies vacuum. The compact instantons are known to be consistent with the vacuum, whereas the Euclidean wormholes may provide additional features. This difference may be distinguished in future observations of, e.g., CMB anisotropies or gravitational waves. In principle the detailed shape of the Euclidean wormhole and the inflation potential can be revealed through the observational signatures. Therefore, future observations may help to impose constraints on the shape of the wormhole and the inflation potential. Specific details of such signatures have to be worked out, which is one of our plans as an extension of the present paper.

There are limitations in our argument. First, the Euclidean path integral may not be the best boundary condition of the Wheeler-DeWitt equation. Second, there may be a significant dependence between the type of instantons ($\mathcal{H}_{c}$ or $\mathcal{H}_{w}$) and the hypothesis that explains a large number of $e$-foldings $\mathcal{H}_{\mathcal{N}}$, while we assumed that there is dependence. However, even if there are such possibilities, we provide an attractive idea to explain various traditional problems of the Euclidean path integral approach.

Our argument is clearly conditional. That is, it is based on the assumption that the Euclidean path integral approach is the correct effective description of the final quantum gravity theory. Before such final theory becomes mature, the persuasion of the Euclidean path integral approach has to rely on its successful applications to other critical issues in quantum gravity or cosmology. For example, if one can calculate any cosmological implications, e.g., CMB observations, from Euclidean wormholes, then it can open a possibility to prove or falsify the quantum gravitational hypothesis based on cosmological observations. We leave these topics for future projects.

\section*{Acknowledgment}
PC is supported by Taiwan National Science Council under Project No.~NSC 97-2112-M-002-026-MY3, Leung Center for Cosmology and Particle Astrophysics (LeCosPA) of National Taiwan University, Taiwan National Center for Theoretical Sciences (NCTS), and US Department of Energy under Contract No. DE-AC03-76SF00515. DR and DY are supported by the Korean Ministry of Education, Science and Technology, Gyeongsangbuk-do and Pohang City for Independent Junior Research Groups at the Asia Pacific Center for Theoretical Physics and the National Research Foundation of Korea (Grant No.: 2018R1D1A1B07049126).

\end{document}